# InP/InAsP Nanowire-based Spatially Separate Absorption and Multiplication Avalanche Photodetectors

Vishal Jain,[1,2] Magnus Heurlin,[1] Enrique Barrigon,[1] Lorenzo Bosco,[1] Ali Nowzari,[1] Shishir Shroff,[1] Virginia Boix,[1] Mohammad Karimi,[1,2] Reza J. Jam,[1] Alexander Berg,[1] Lars Samuelson,[1] Magnus T. Borgström,[1] Federico Capasso[3] and Håkan Pettersson[1,2*]

[1]Solid State Physics and NanoLund, Lund University, Box 118, SE-221 00 Lund, Sweden

[2]Laboratory of Mathematics, Physics and Electrical Engineering, Halmstad University, Box 823, SE-301 18 Halmstad, Sweden

[3]School of Engineering and Applied Sciences, Harvard University, Cambridge, Massachusetts 02138, USA

*Corresponding author e-mail: hakan.pettersson@hh.se

ABSTRACT

Avalanche photodetectors (APDs) are key components in optical communication systems due to their increased photocurrent gain and short response time as compared to conventional photodetectors. A detector design where the multiplication region is implemented in a large bandgap material is desired to avoid detrimental Zener tunneling leakage currents, a concern otherwise in smaller bandgap materials required for absorption at 1.3/1.55 μm. Self-assembled III-V semiconductor nanowires offer key advantages such as enhanced absorption due to optical resonance effects, strain-relaxed heterostructures and compatibility with main-stream silicon technology. Here, we present electrical and optical characteristics of single InP and InP/InAsP nanowire APD structures. Temperature-dependent breakdown characteristics of $p^+$-n-$n^+$ InP nanowire devices were investigated first. A clear trap-induced shift in breakdown voltage was inferred from I-V measurements. An improved contact formation to the $p^+$-InP segment was observed upon annealing, and its effect on breakdown characteristics was investigated. The bandgap in the absorption region was subsequently varied from pure InP to InAsP to realize spatially separate absorption and multiplication APDs in heterostructure nanowires. In contrast to the homojunction APDs, no trap-induced shifts were observed for the heterostructure APDs. A gain of 12 was demonstrated for selective optical excitation of the InAsP segment. Additional electron beam-induced current measurements were carried out to investigate the effect of local excitation along the nanowire on the I-V characteristics. Our results provide important insight for optimization of avalanche photodetector devices based on III-V nanowires.

The research timeline of III-V semiconductor-based avalanche photodetectors (APDs) spans over half a century.[1, 2] More recently, improved gain and sensitivity have been attained by shrinking the impact-ionization region in APDs based on nanostructures[3] and nanowires (NWs).[4, 5] Furthermore, APDs with absorption in the commercially desired 1.0-1.6 μm wavelength region have also been reported by incorporating quantum heterostructures within the high electric-field region.[6, 7] None of these reported devices utilize the strong nanophotonic resonances exhibited by properly designed NW arrays for efficient light absorption.[8-11] One approach is to use a plasmonic optical antenna to physically separate the absorption and multiplication regions at the tip and base of NWs, respectively.[12] When a low bandgap material is used for the multiplication region, the devices typically suffer from significant Zener tunneling leakage currents. Therefore, a separate absorption and multiplication (SAM) APD heterojunction structure was proposed four decades ago.[13, 14] This motivated us to develop and now report on the design and realization of the first heterostructure SAM APD implemented in a single NW to this date. In these APDs the absorption and multiplication regions are implemented in InAsP, spectrally tuned at 1.55 μm, and InP, respectively. To eliminate possible effects of complex processing in NW-arrays, this study entirely focuses on laterally contacted single NW APDs. Our results pave the way for future realization of NW-based array SAM APDs compatible with main-stream silicon technology.

First, a simpler homojunction structure of $p^+$-n-$n^+$ InP NWs (sample A) was grown to investigate the breakdown characteristics. This was followed by incorporation of an $n^-$-InAsP absorption segment forming a ($p^+$-n)-InP-$n^-$-InAsP-$n^+$-InP NW SAM APD (sample B) structure. To obtain statistics from a uniform set of NWs, nanoimprint lithography (NIL) was employed where periodic patterns of circular holes with a diameter of 180 nm and a pitch of 400 nm were prepared on p$^+$-InP (111)B substrates followed by metal evaporation and lift-off of 20 nm Au film.[15] An alternative approach to define the gold seed particles in nano-sized holes would be to use our recently developed Au electrodeposition technique.[16] NW growth was then carried out in a low-pressure (100 mbar) metal organic vapor phase epitaxy (MOVPE) system (Aixtron 200/4), with a total flow of 13 l/min using hydrogen (H$_2$) as carrier gas. Trimethylindium (TMI) , arsine (AsH$_3$) and phosphine (PH$_3$) were used as precursors for NW growth while diethylzinc (DEZn) and tetraethyltin (TESn) were used as p-dopant[17] and n-dopant[18] precursors, respectively. Hydrogen chloride (HCl) at a molar fraction of $\chi_{HCl} = 6.1 \times 10^{-5}$ was used to inhibit the radial growth.[19] Before growth, the samples were first annealed at 550 ºC for 10 min under a PH$_3$/H$_2$ gas mixture to desorb any surface oxides at a constant molar fraction of $\chi_{PH3} = 6.9 \times 10^{-3}$. In this step the Au discs defined by NIL turn into 130 nm diameter catalyst nanoparticles. The reactor was then cooled to 440 ºC, at which point growth was initiated by the addition of TMI and DEZn to the gas flow at molar fractions of $\chi_{TMI} = 7.4 \times 10^{-5}$ and $\chi_{DEZn} = 6.1 \times 10^{-5}$ and lowering $\chi_{PH3}$ to $1.5 \times 10^{-3}$. After a 30 s nucleation time, HCl was introduced and $\chi_{PH3}$ was increased to $6.9 \times 10^{-3}$. The grown InP p$^+$-segment has a nominal acceptor concentration of about $5 \times 10^{18}$ cm$^{-3}$.[17] For the n-segment, TESn was used at a molar fraction of $\chi_{TESn} = 3.5 \times 10^{-7}$ resulting in an estimated doping concentration of $1$-$5 \times 10^{16}$ cm$^{-3}$ based on previous studies.[18, 20, 21] For the top n$^+$-segment growth, $\chi_{TESn} = 4.8 \times 10^{-6}$ along with a reduced TMI flow rate ($\chi_{TMI} = 5.2 \times 10^{-5}$) was used to effectively increase the Sn incorporation[22] to an estimated doping level higher than $5 \times 10^{18}$ cm$^{-3}$.[21] After completing the top n$^+$-segment, the growth was terminated and the sample cooled down in a PH$_3$/H$_2$ gas mixture. The grown p$^+$-n-n$^+$ InP NWs had a diameter of 130 nm and average segment lengths of approximately 500/1100/350 nm for a growth time of 2/6/9 min, respectively, measured using an in-situ optical reflectance characterization setup.[23] For electrical measurements, single NWs were mechanically transferred to n-type Si substrates coated with 100 nm thermal $SiO_2$ and 12 nm $HfO_2$. The samples were spin-coated with PMMA 950A5 at 5000 rpm followed by a 5 min hot plate bake at 180 °C. The contact regions to the NW ends were exposed using a Raith 150 electron beam lithography system, with subsequent pattern development in a MIBK:IPA 1:3 solution for 1 min. Then a multilayer of Pd/Zn/Pd/Au was thermally evaporated with thicknesses of 2/20/68/110 nm, respectively, and removed from unexposed areas through lift-off in acetone. Figure 1a shows an SEM image of a fabricated device. The I-V characteristics of the contacted NWs were measured with a Keithley 6430 sub-femtoampere sourcemeter with the samples mounted inside a Janis PTSHI-950-FTIR pulse-tube closed-cycle cryostat integrated with a Bruker Vertex 80V Fourier transform infrared (FTIR) spectrometer. The built-in quartz lamp in the FTIR was used as the light source for recording the I-V under illumination.

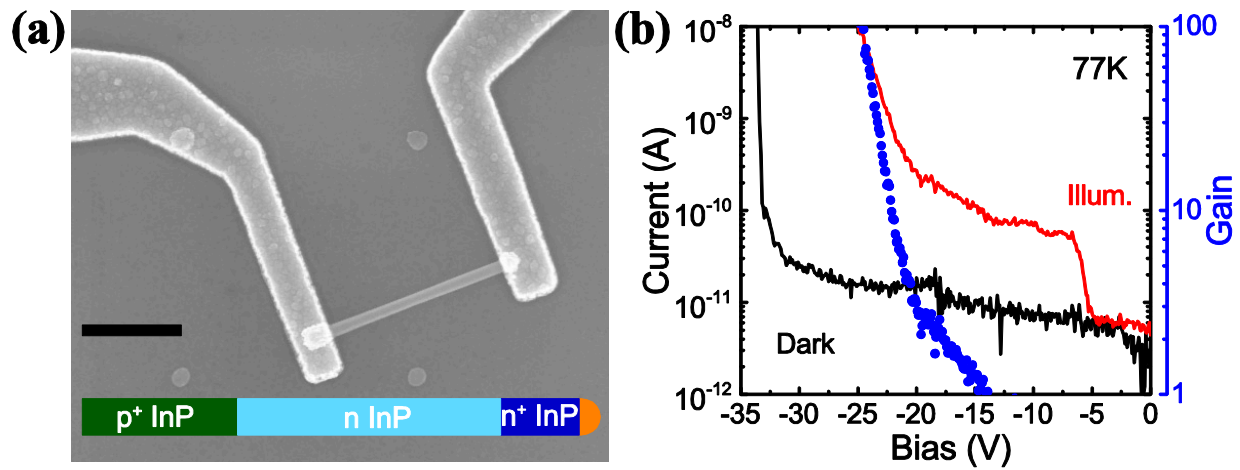

**Figure 1.** (a) SEM image of a laterally-contacted $p^+$-n-$n^+$ InP NW device along with the schematics. The scale bar is 1 μm. (b) Avalanche breakdown characteristics of a corresponding device at 77 K under darkness and illumination with quartz-lamp, plotted together with the gain calculated from the I-V characteristics.

A low dark current (~10 pA) was recorded for sample A at 77 K before the onset of a breakdown with 170 mV/decade current rise at a reverse bias of about 33 V (Figure 1b). Several individually contacted single NW-devices were studied where the breakdown voltage was found to vary in a range of 31±6 V at 77 K. This range could be an effect of doping variations between different NWs. The breakdown voltage has a positive temperature dependence (Figure 2a) with an estimated temperature coefficient of 50±20 mV/K (measured up to 150 K, clear breakdown characteristics were not observed at higher temperatures) for different devices, which is in agreement with the reported values for bulk InP.[24] Under illumination with a quartz-lamp at 77 K, we observe the onset of a punch-though region at about 5 V where the photocurrent increases by a factor of 10 (Fig. 1b). Typically, punch-through is observed when the depletion region extends to the edge of the absorbing layer in a SAM APD. In this case of InP APDs, the relatively high doping in the absorption region (1-5×$10^{16}$ cm$^{-3}$) lead to recombination of the photogenerated carriers before they could be collected at low biases. At about 5 V reverse bias, the depletion region extends close enough to the n$^+$-segment (Supporting Information Figure S1, in agreement with expected segment lengths estimated from growth parameters and SEM images) to facilitate an efficient collection of photogenerated carriers. As the punch-through occurs before the field in the multiplication region is sufficiently high to produce avalanche gain, the relatively flat photocurrent above punch-through (from 5-15 V) can be used as the unity gain (UG) reference point.[25] Using the McIntyre's equation[26] for an estimation of the gain $G = (I_{ill} - I_{dark})/(I_{ill}(UG) - I_{dark}(UG))$, we observe a gain factor of up to 80 at 25 V in the best devices. We note that the breakdown voltage of about 20V under illumination is significantly lower than under dark conditions. A similar voltage shift is in fact typically observed in darkness when the I-V characteristic is re-swept after the initial sweep (Fig. 2a). Subsequent sweeps, however, do not result in any further significant shifts and the I-V characteristics remain constant (Supporting Information Figure S2). We attribute the observed voltage shift to filling of hole-traps after the first sweep, thereby increasing the effective space charge density and thus reducing the bias required to obtain the high electric field necessary for breakdown. A similar effect has recently been reported in the case of SiC avalanche diodes.[27] The effect of traps is also clearly visible under power-dependent laser excitation (532 nm) where the photocurrent saturates at higher laser power (Figure 2b). Further time-dependent studies, beyond the scope of this work, are needed to examine the trapping mechanisms in detail.

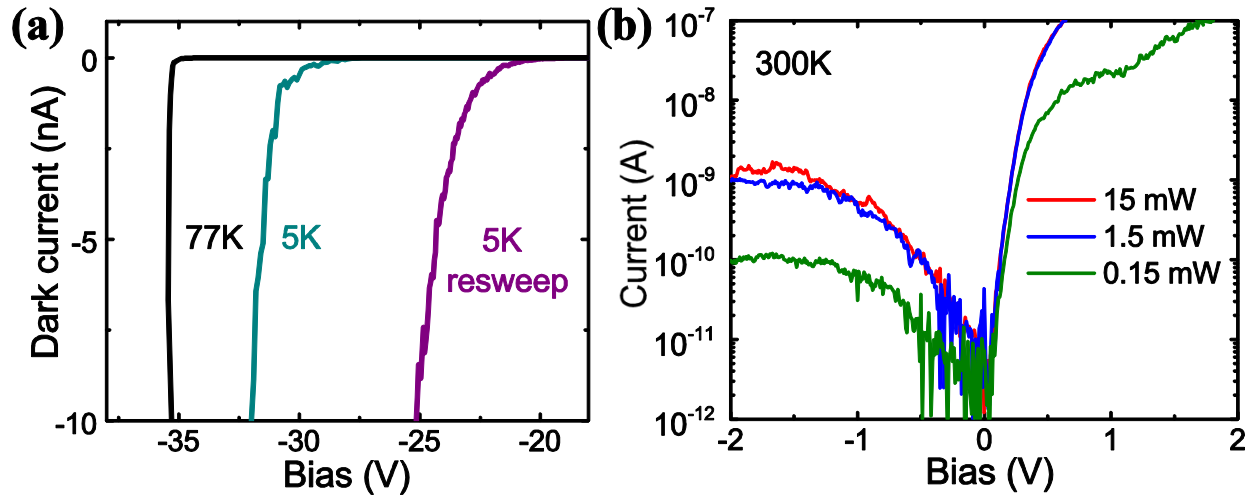

**Figure 2.** I-V characteristics of a typical device from sample A. (a) Temperature dependent shift of the breakdown voltage. The first re-sweep of the bias shows a significant shift due to trapped carriers. (b) I-V characteristics under illumination with a 532 nm laser with increasing power.

An interesting observation is the effect of wafer dimensions on the breakdown characteristics. For each growth run, we placed two Au catalyst imprinted samples of different size in the MOVPE chamber. It was found that the NWs obtained from the smaller sample showed a much higher breakdown voltage (67±16 V at 77 K, also shown in Figure 3b at 150 K) compared to the NWs discussed above (31±6 V at 77 K). We attribute this observation to a larger edge effect for the smaller sample, which results in longer NWs (2.2-2.4 µm in this case compared to 1.6-1.8 µm in the previous set) which in turn is a result of more In being present close to the sample edges. A higher amount of In reduces the Sn dopant incorporation[22] and thereby increases the bias required to obtain sufficiently high electric fields for impact-ionization.

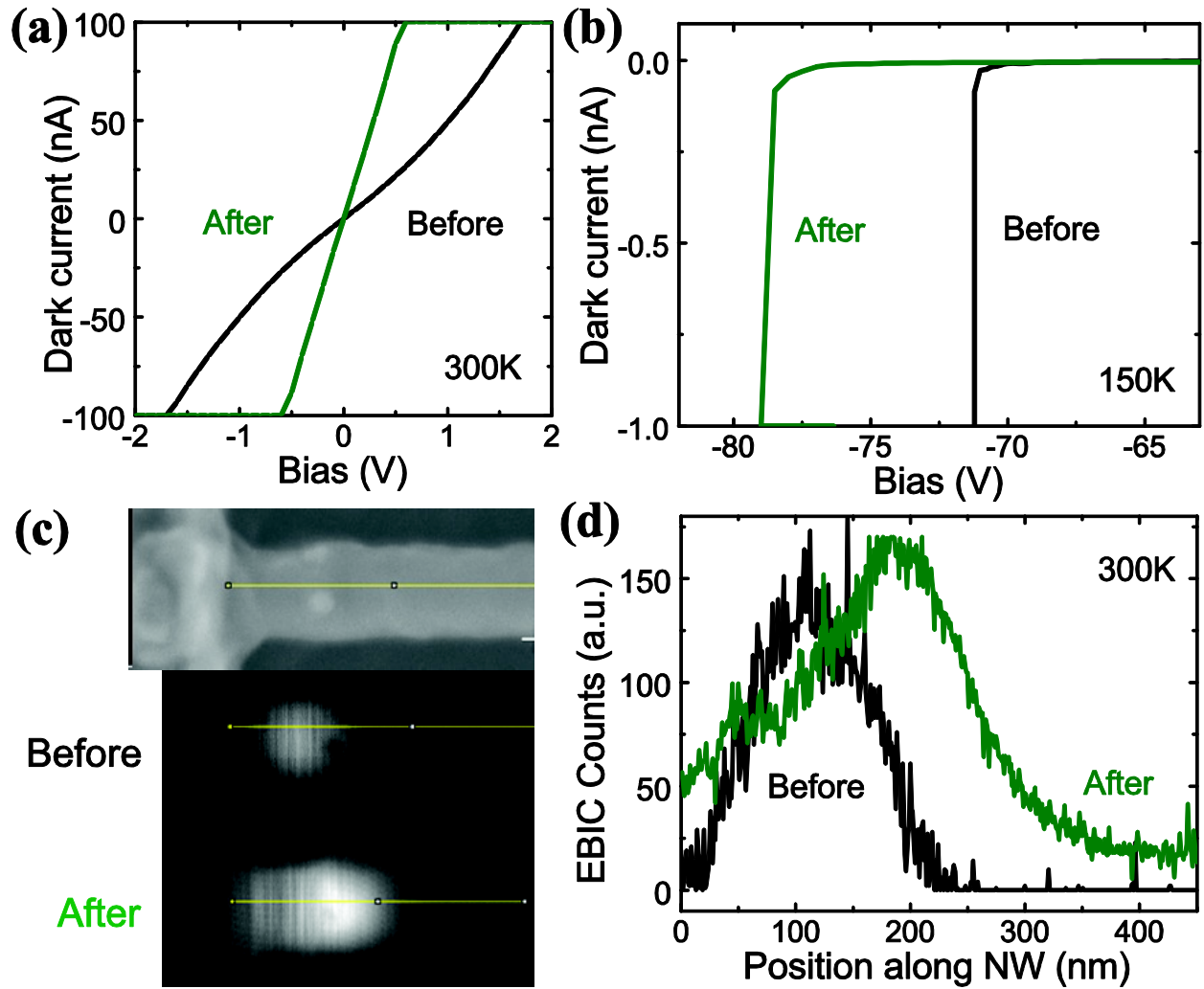

**Figure 3.** (a) Improved contact resistance between two metal contacts on a $p^+$-segment after annealing. (b) Breakdown characteristics before and after annealing at 350 °C of a $p^+$-n-$n^+$ InP NW APD. (c) SEM image along with EBIC signal and (d) measured EBIC profile along the NW before and after annealing.

It is well known that ohmic contacts are difficult to obtain to $p^+$-InP.[28] It has been demonstrated that an annealing step above 420 °C allows Zn from the metal contact to diffuse into the InP NWs which significantly improves the contact properties.[29] This approach has previously also been used for $p^+$-InP NWs.[30] Annealing above 350 °C, however, led to changes in surface morphology in our NW devices (Supporting Information Figure S3). The contact to the $p^+$-InP segment in our devices was thus activated by a rapid thermal annealing process at 350 °C for 30 s. We measured a substantial decrease in contact resistance between two nearby Pd/Zn/Pd/Au contacts to the $p^+$-InP segment after annealing (Figure 3a). This contact resistance effect was further studied using $e^-$-beam induced current (EBIC) measurements. Before annealing, the EBIC signal from the Schottky contact between the Pd/Zn/Pd/Au contact and $p^+$-InP segment was dominant, whereas the effective $p^+$-n junction in the NW was clearly visible after annealing (Figure 3c, d). The $p^+$-n junction boundary observed at a distance of 200 nm from the metal contact (Figure 3c) is reasonable considering that out of the 500 nm $p^+$-InP segment, about 190 nm is buried under the metal contact and about 60-90 nm of the NW base is left as a stub on the as-grown substrate after transferring the NWs (Supporting Information Figure S4). The annealing step further led to an increase in the observed breakdown voltage (Figure 3b). We can only speculate that the diffused Zn might act as a compensatory dopant in the adjacent n-InP segment, thus requiring a higher applied bias to achieve the necessary electric field.

After the investigation of the homojunction InP APDs, heterostructure APDs (sample B) were realized by changing the material in the absorption region from InP to InAsP by tuning the As composition. The same growth parameters as for sample A were used to grow a 500 nm (2 min) $p^+$-InP segment and a 400 nm (2 min) n-InP segment. To maintain an almost constant As concentration in the 700 nm (8 min) $n^-$-InAsP segment the AsH$_3$ flow was ramped from a molar fraction of $\chi_{AsH3} = 2.3\times10^{-4}$ to $\chi_{AsH3} = 1.2\times10^{-4}$. To compensate for the lower etch rate of InAsP as compared to InP the HCl flow was increased to a molar fraction of $\chi_{HCl} = 6.1\times10^{-5}$, thereby reducing tapering of the NWs. A reduced flow of TESn at a molar fraction of $\chi_{TESn} = 3.5\times10^{-8}$ resulted in an estimated doping concentration of 5-8×10$^{15}$ cm$^{-3}$.[18, 20] The gas flows were finally switched back to the same parameters used for the growth of the top n$^+$-InP segment in sample A, but resulting in a shorter 140 nm long n$^+$-InP segment. From TEM investigations it was concluded that the grown InAsP segment is predominantly wurtzite with an almost constant As concentration of 63 ± 2 %. We also found a 100 nm long segment next to the n$^-$-InP segment where the As concentration was 53±2 % (Figure 4a). Although this segment was not intended, it can increase the efficiency of hole extraction from the absorption segment since the valence band discontinuity between the absorption and multiplication segment is reduced.[31]

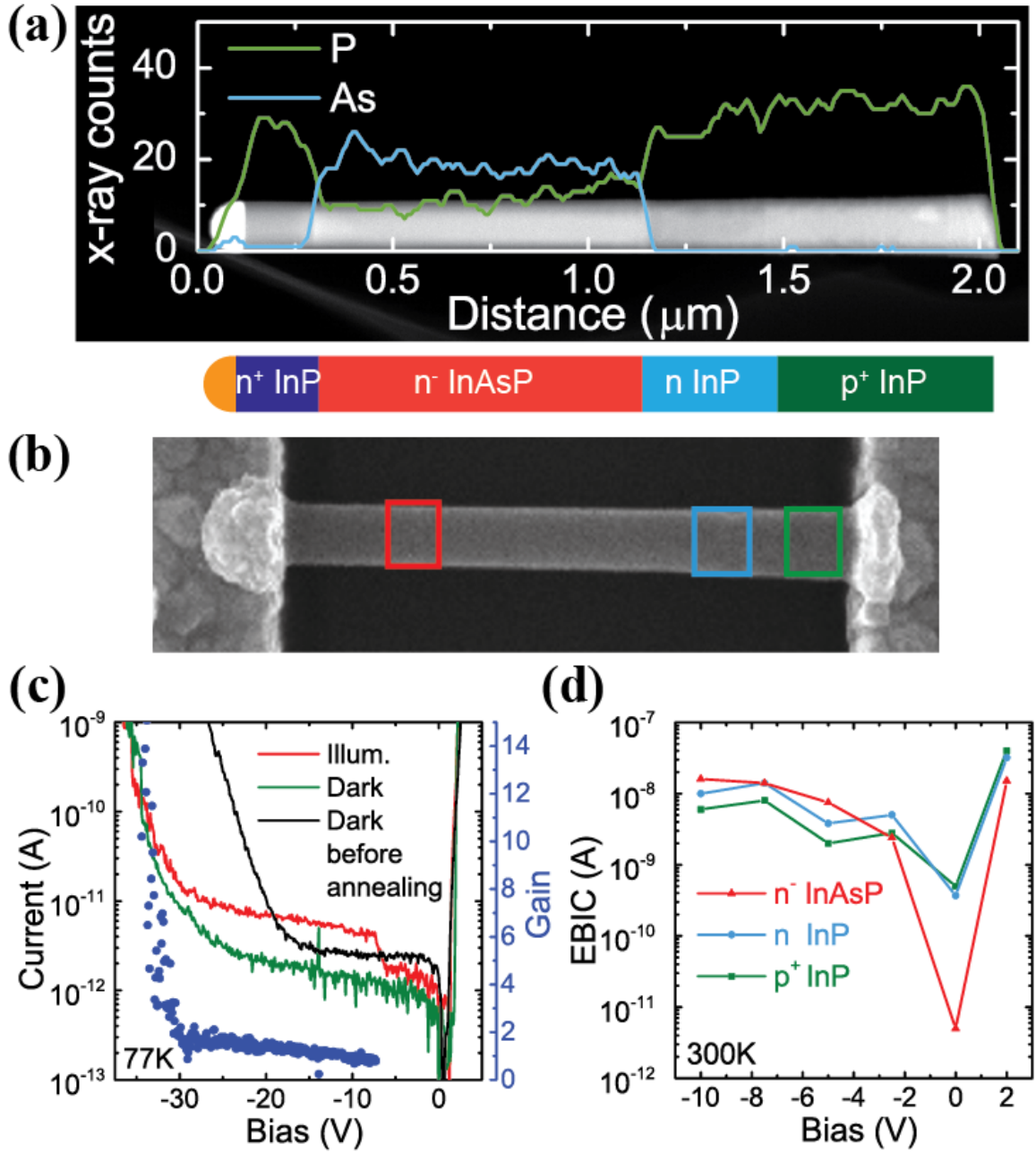

**Figure 4.** (a) EDX linescans superimposed on a TEM image, along with a schematics of an InP/InAsP NW SAM APD (sample B). (b) Spatially resolved e-beam excitation map over the length of the NW. (c) I-V characteristics and gain at 77 K of an InP/InAsP NW SAM APD in darkness, before and after annealing, and under selective illumination of the InAsP absorption region. (d) I-V characteristics corresponding to e-beam excitation in (b).

The grown NWs were transferred to Au-patterned Si substrates for photoluminescence (PL) measurements at 4 K. PL signals corresponding to the InP regions and n$^-$-InAs$_{0.63}$P$_{0.37}$ absorption region were observed at about 0.90 μm and 1.55 μm at 4 K (Supporting Information Figure S5). In addition we measured a significant PL signal in the 1.2-1.4 μm range, the origin of which is unclear at this point. A comparative PL spectrum from an InP APD shows no such peaks which rules out the possibility of dopant-related defects. The rather sharp peaks most likely originate from defects at the InP-InAsP interface. Further studies using characterization techniques like cathodoluminescence spectroscopy are needed to unravel the origin of these peaks. We point out, however, that these defects do not seem to affect the device performance as evident from the discussions below.

The contacted SAM APDs in general showed less sharp breakdown characteristics (Figure 4c) compared to InP APDs, even at 77 K. This might be attributed to an increased tunnelling leakage current due to the smaller bandgap ($E_g$ = 0.8 eV) of InAsP, in spite of our efforts to tailor the electric field distribution (doping profile) to suppress such breakdown mechanisms.[14] Interestingly, these devices exhibited minimal shift in I-V characteristics in subsequent sweeps (Supporting Information Figure S6), indicating that the previously observed trapping effects are associated with InP. They do, however, show similar breakdown voltage shifts upon annealing (Figure 4c) as the InP APDs. For illumination, a quartz lamp with a silicon long-pass filter was used to selectively illuminate the InAsP absorption region. At the punch-through bias of 8 V, the photocurrent increases by a factor of 4 to a unity gain current of 4 pA yielding a gain factor of up to 12. It is worth mentioning here that a quartz lamp was used in these experiments instead of a high-power laser to demonstrate the usability of the NW-based APDs. Almost no increase in total current was observed upon illumination with a quartz lamp without the silicon long-pass filter for reverse biases below the punch-through region (Supporting Information Figure S7), which indicates that the carriers are selectively generated and collected from the InAsP absorption region.

A further verification was done through selective spatial excitation with an e$^-$-beam (Figure 4b). Exciting the p$^+$-InP or n-InP region yielded similar low current levels at zero or forward biases, but with a slightly enhanced drift-induced carrier collection at reverse biases when bombarding the n-InP region with electrons (Figure 4d). Exciting the n$^-$-InAsP region generated significantly lower current levels at zero bias due to the InP/InAsP heterojunction barrier faced by the holes, which is more easily overcome at forward bias. Interestingly, the EBIC signal increased by a factor of 4 at 10 V with respect to 2 V reverse bias, similar to the photocurrent increase under punch-through conditions.

In conclusion, we have studied avalanche breakdown characteristics of single NW InP APDs where a multiplication gain factor of up to 80 was observed. Annealing the samples at 350 °C significantly improved the contact to p$^+$-InP and increased the breakdown voltage. A clear effect of trap states on the breakdown characteristics was evident and a detailed study is needed to investigate the trap mechanism. The main result of the work is the first demonstration of InP/InAsP NW SAM APDs with spectrally tuned absorption at 1.55 μm. Studies of selective excitation of the absorption region using both photons and electrons yielded consistent results and a multiplication gain of up to 12. These devices are ideal candidates for future development of NW-based array SAM APDs.

ASSOCIATED CONTENT

**Supporting Information**

This material is available free of charge via the Internet at http://pubs.acs.org.

AUTHOR INFORMATION

**Corresponding Author**

* hakan.pettersson@hh.se

All authors have given approval to the final version of the manuscript.

ACKNOWLEDGEMENTS

The authors acknowledge financial support from NanoLund, the Swedish Research Council, the Swedish National Board for Industrial and Technological Development, the Swedish Foundation for Strategic Research, the Ljungberg Foundation, the Carl Trygger Foundation and the Swedish Energy Agency. This project has received funding from the European Union's Horizon 2020 research and innovation program under grant agreement No 641023 (NanoTandem) and under the Marie Sklodowska-Curie grant agreement No 656208.

Supporting Information for

# InP/InAsP Nanowire-based Spatially Separate Absorption and Multiplication Avalanche Photodetectors

*Vishal Jain,[1,2] Magnus Heurlin,[1] Enrique Barrigon,[1] Lorenzo Bosco,[1] Ali Nowzari,[1] Shishir Shroff,[1] Virginia Boix,[1] Mohammad Karimi,[1,2] Reza J. Jam,[1] Alexander Berg,[1] Lars Samuelson,[1] Magnus T. Borgström,[1] Federico Capasso[3] and Håkan Pettersson[1,2]\**

[1]Solid State Physics and NanoLund, Lund University, Box 118, SE-221 00 Lund, Sweden

[2]Laboratory of Mathematics, Physics and Electrical Engineering, Halmstad University, Box 823, SE-301 18 Halmstad, Sweden

[3]School of Engineering and Applied Sciences, Harvard University, Cambridge, Massachusetts 02138, USA

*Corresponding author e-mail: hakan.pettersson@hh.se

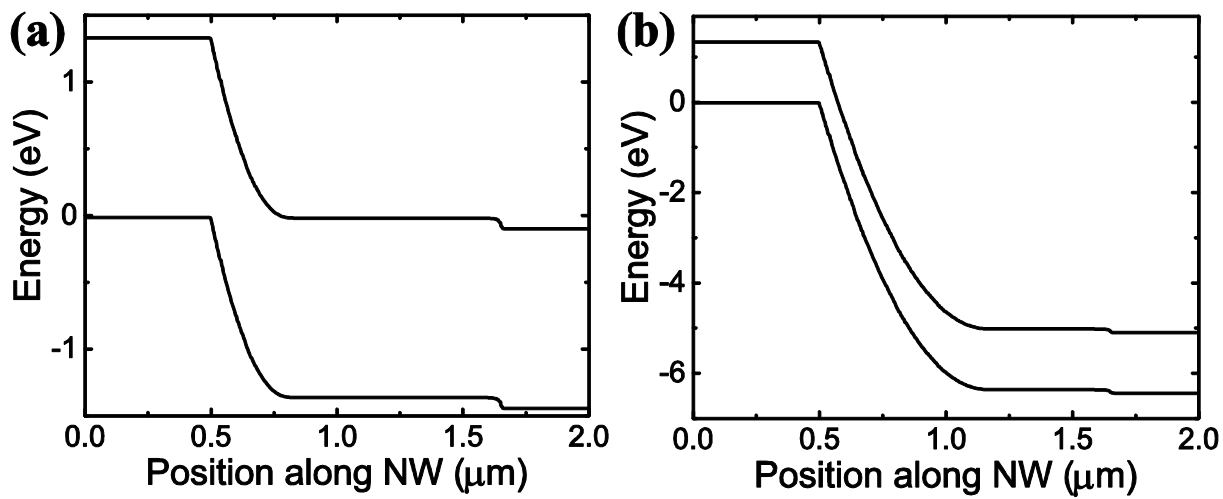

**Figure S1.** Energy band diagram at 77 K at (a) 0 V and (b) 5 V reverse bias of a $p^+$-n-$n^+$ InP NW APD (sample A).

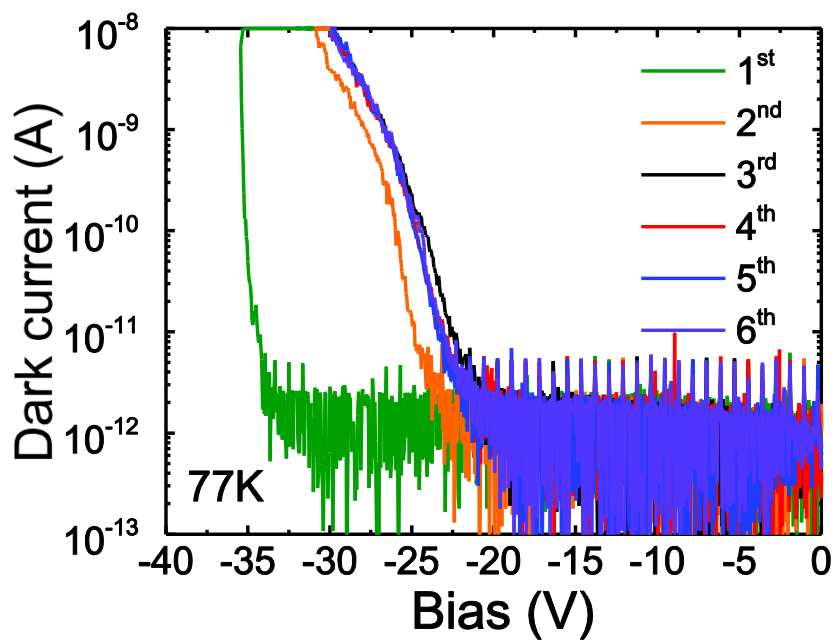

**Figure S2.** Breakdown voltage shift upon successive sweeps for sample A.

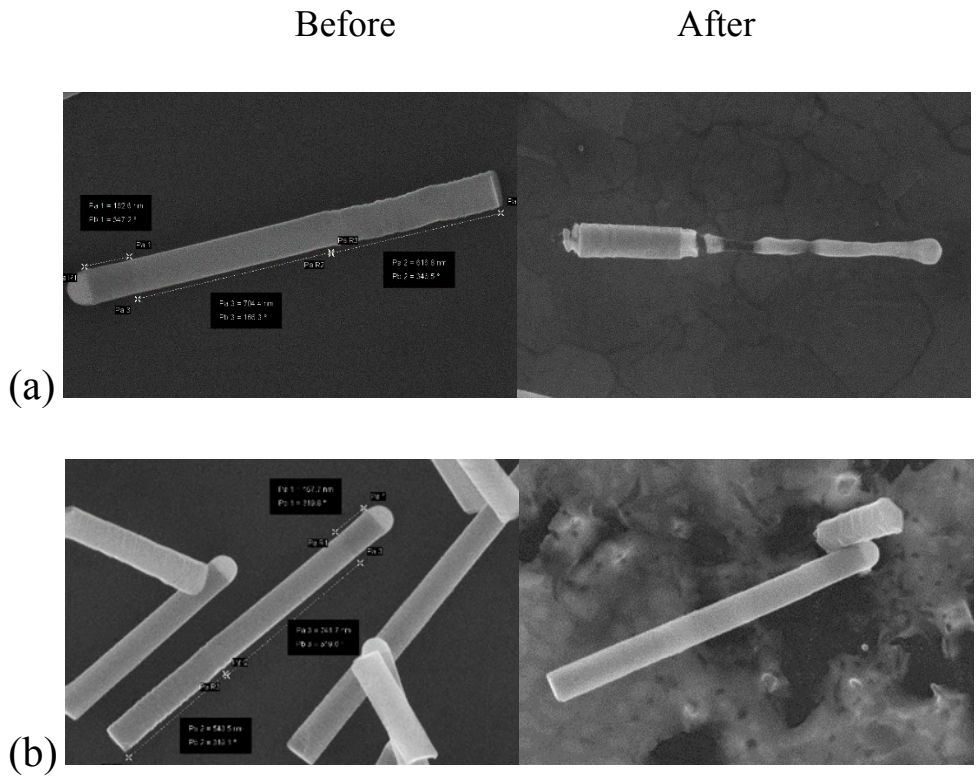

**Figure S3.** Changes in NW surface morphology after annealing at (a) 400 °C and (b) 375 °C.

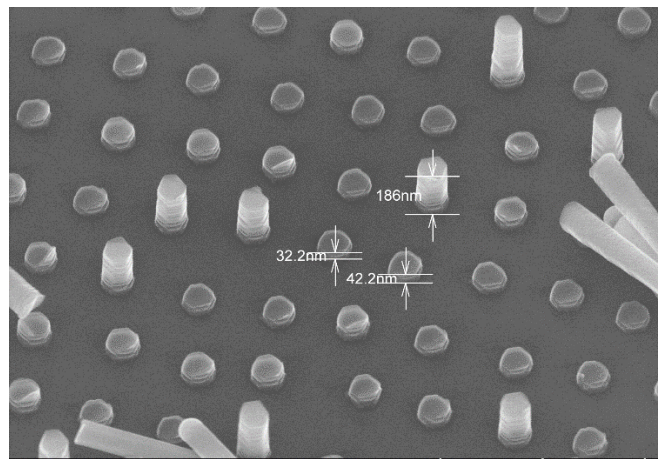

**Figure S4.** NW stubs remaining on the as-grown substrate after NW transfer.

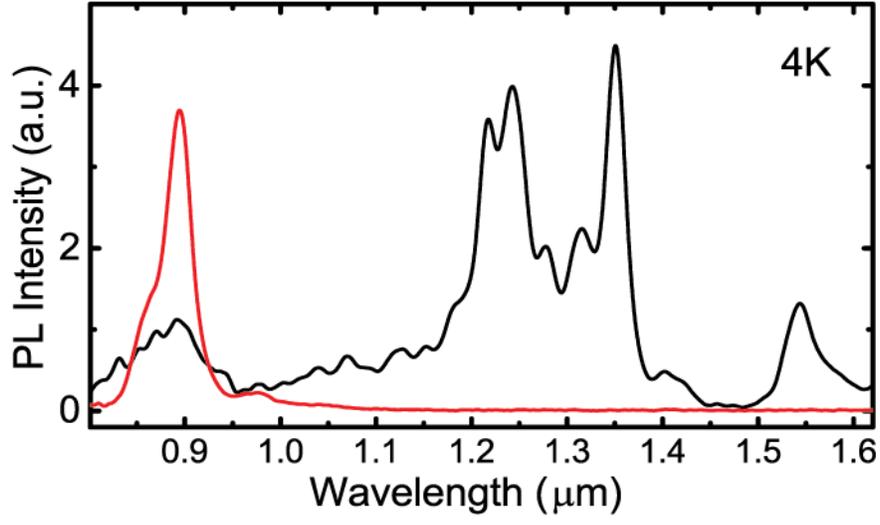

**Figure S5.** PL spectrum of a representative single InP/InAsP SAM NW APD (sample B, black trace), compared with the corresponding PL spectrum of an InP NW APD (sample A, red trace).

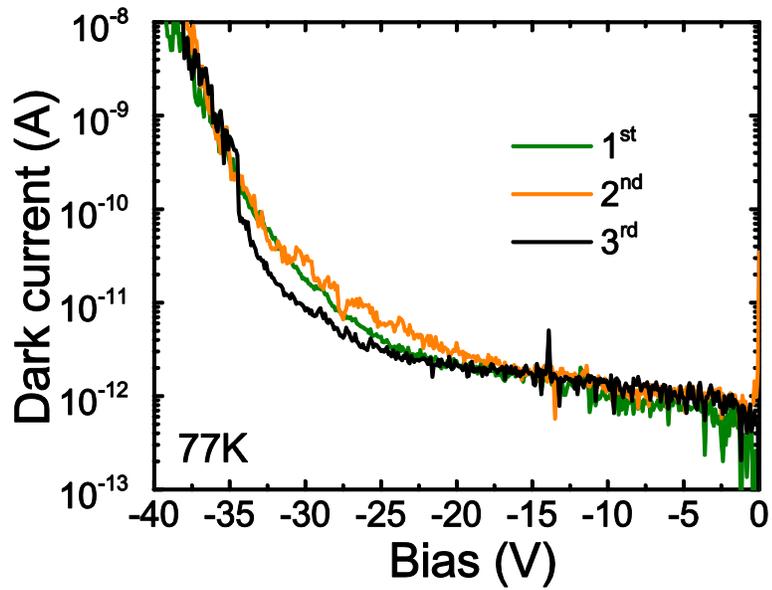

**Figure S6.** I-V characteristics upon successive sweeps for sample B.

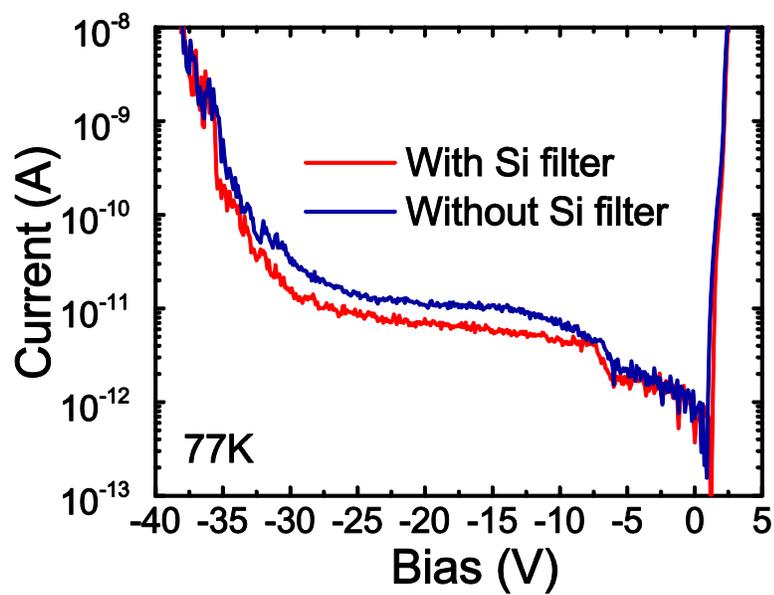

**Figure S7.** Comparison of I-V characteristics of sample B under illumination with a quartz lamp with and without a long-pass silicon filter.